\documentclass[twocolumn,superscriptaddress,floatfix,prl,aps,showpacs]{revtex4}
\usepackage{graphicx}
\usepackage{latexsym}
\usepackage{amsmath}
\begin{document}
\bibliographystyle{apsrev}
\title{Knight shifts around vacancies in the 2D Heisenberg model}
\author{Fabrizio Anfuso}
\affiliation{Institute of Theoretical Physics,
Chalmers University of Technology and G\"oteborg University,
S-412 96 G\"oteborg, Sweden}
\author{Sebastian Eggert}
\affiliation{Institute of Theoretical Physics,
Chalmers University of Technology and G\"oteborg University,
S-412 96 G\"oteborg, Sweden}
\affiliation{Dept.~of Physics, Univ.~of Kaiserslautern, Erwin-Schr\"odinger Str., D-67663 Kaiserslautern, Germany}
\begin{abstract}
The local response to a uniform field around vacancies in the two-dimensional (2D)  spin-1/2
Heisenberg antiferromagnet is determined by numerical quantum Monte Carlo simulations
as a function of temperature.
It is possible to separate the Knight shifts into uniform and staggered
contributions on the lattice which are analyzed and understood in detail.
The contributions show interesting long and short range behavior that may be of relevance in
NMR and susceptibility measurements.  For more than one impurity remarkable non-linear enhancement 
and cancellation effects take place. We predict that the 
Curie impurity susceptibility will be observable 
for a random impurity concentration even in the thermodynamic limit.
\end{abstract}
\maketitle

The deliberate use of substitutional impurities in strongly correlated electron systems has 
become a valuable tool 
for controlled studies of the underlying correlations\cite{Xiao,Mahajan,Ting,Corti,Vajk}.  
Doping of non-magnetic Zn-ions in antiferromagnetic CuO$_2$ planes is known to lead
to a surprisingly large reduction of $T_c$\cite{Xiao} and induces 
local staggered magnetic moments around the impurity 
sites\cite{Mahajan}.  It has now become quite standard to analyze the local magnetic moments around
static magnetic and non-magnetic impurities in an antiferromagnetic background for a 
deeper understanding of the correlated states\cite{Ting,Corti,Vajk}.

Theoretically the Knight shifts around impurities and vacancies have been studied for many 
low-dimensional antiferromagnets\cite{Bulut,eggert,Martins,Sandvik2,Bulut1,rommer,eggert2},
which typically show
a strong enhancement of the antiferromagnetic order.
The detailed behavior of the 
staggered magnetic moments can even give rather exotic results
such as an increasing amplitude as a function of distance\cite{eggert} and show renormalization 
effects\cite{eggert2,rommer}. 
The magnetization pattern can often be 
interpreted as the interference of incoming and scattered quasiparticle excitations\cite{rommer} or
in terms of a pruned valence bond basis\cite{Martins}.

The sum of all Knight shifts adds up to the total 
susceptibility $\chi_1$.  The resulting impurity susceptibility $\chi_{\rm imp} = 
\chi_1 - \chi_0$ has been studied in detail in more recent theoretical 
studies\cite{Sachdev,Nagaosa,Sandvik,Sandvik1,Sachdev2,Sachdev3,Sushkov,Sachdev1}.
In systems with long range magnetic order the 
leading temperature dependence of the impurity susceptibility from one single vacancy is given by a 
{\it classical} Curie spin which has 
been confirmed by numerical simulations for the 2D  Heisenberg antiferromagnet\cite{Sandvik,Sandvik1}, 
where 
a subleading logarithmic term has also been established\cite{Sandvik,Sachdev,Sachdev1,Sushkov,Sandvik1}
\begin{equation}
\chi_{\rm imp}=\chi_1-\chi_0=\frac{S^2}{3T}+\frac{1}{3 \pi \rho_s}\ln\left(\frac{C}{T}\right). \label{chiimp}
\end{equation}
Here $\chi_0$ is the total susceptibility without any impurities, $\rho_s$ is the spin stiffness,  
and the limit of large correlation length $\xi(T)$ and system
size $\xi(T)> L \to \infty$ is assumed. 

We now want to examine how this impurity susceptibility is distributed on the lattice by considering
the linear {\it local} response 
to a small uniform magnetic field (Knight shift)
at each site 
\begin{equation}
\chi(\mathbf{r})=\beta\sum_i\langle S^z_i S^z_{\mathbf{r}} \rangle.
\end{equation}
We find that long and short range patterns of the Knight shifts contribute to the 
impurity susceptibility differently.  We are able to study the interference
of several impurities with non-trivial cancellation and enhancement effects over long distances, 
which allows us to make predictions for a finite impurity density.

In general we can write $\chi(\mathbf{r})$ as a sum of uniform and staggered parts
on the lattice
\begin{equation}
\chi(\mathbf{r}) = \chi_{\rm uni}(\mathbf{r})+(-1)^{r_x+r_y}\chi_{\rm stag}(\mathbf{r}),
\end{equation}
the amplitudes of which are both slowly varying on the scale of one lattice spacing.  
In order to extract those two 
components we extrapolate the data on the even sublattice to the odd sublattice and 
vice versa and define
 $\chi_{\rm uni/stag}(\mathbf{r}) =(\chi_{\rm even}(\mathbf{r})\pm\chi_{\rm odd}(\mathbf{r}))/2$.
The model is the
2D Heisenberg spin-1/2 antiferromagnet 
$H=J\sum_{\langle i,j \rangle}\mathbf{S}_{i}\cdot \mathbf{S}_{j}$
where $\langle i,j \rangle$ denotes nearest neighbor  sites on a periodic square lattice.
The quantum Monte Carlo program we developed uses the loop algorithm in a 
single cluster variety implemented in continuous
time\cite{Evertz,Evertz1,Wolff,Beard}, which gives efficient updates even at very 
low temperatures on a lattice of $100\times 100$ sites.
For convenience temperatures and energies are given in units of $J=1$.

\begin{figure}
\includegraphics[width=.47\textwidth]{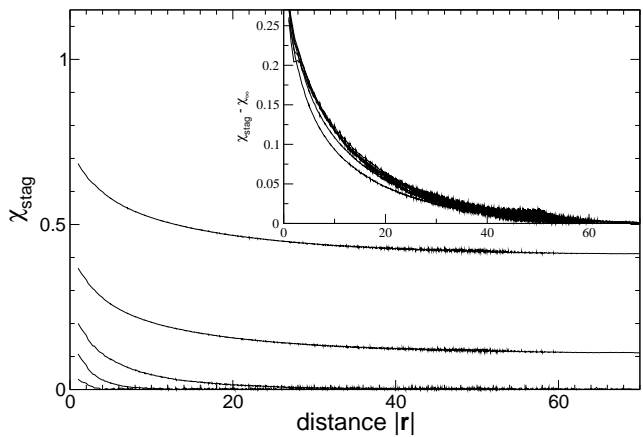}
\caption{All data points of $\chi_{\rm stag}(\mathbf{r})$ as a function of 
distance $|\mathbf{r}|$ from one vacancy for $T=1,0.5,0.33,0.2,0.1$ from below. 
Inset:  All curves collapse to one almost universal, temperature independent shape after subtracting the 
limiting constant of the induced order $\chi_\infty$ for $0.02\leq T\leq 0.2$. }
\label{diagonal}
\end{figure}
\begin{figure}
\includegraphics[width=.47\textwidth]{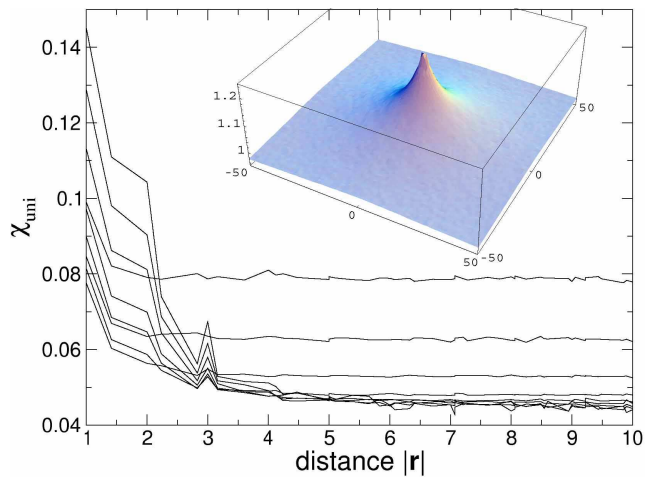}
\caption{(Color online) The uniform part $\chi_{\rm uni}(\mathbf{r})$ as a function of 
distance $|\mathbf{r}|$ for $T=0.5,0.33,0.2,0.1,0.067,0.05, 0.033, 0.025, 0.02$ 
(from above at $\mathbf{r}=10$). Inset: The staggered part at $T=0.05J$}
\label{unipl}
\end{figure}

The Knight shift data around one single vacancy is remarkably isotropic as a function of geometrical 
distance at all temperatures as is shown in Fig.~\ref{diagonal} and the inset 
of Fig.~\ref{unipl} for the staggered part. 
Slight anisotropic deviations near the edges ($|\mathbf{r}| \sim 40 - 50$) can be seen 
due to the finite system size and periodic boundary conditions.
The remarkable isotropy on an anisotropic lattice is an indication that the observed behavior is likely
due to spin wave excitations, which have an isotropic dispersion at low energies.
At high temperatures $T$ the impurity effect is confined within the correlation length $\xi(T)$,
which has an exponential dependence with $1/T$\cite{Chakravarty,Hasenfratz,Beard1}.
Already for $\beta=5$ the impurity affects the whole system in our case. 
This is commonly referred to as the zero temperature limit where N\'eel order effects 
can be observed in 2D. 
In finite systems the effects of this order can be observed already for finite
temperatures when the correlation length exceeds the system size $\xi(T)>L$ (in our case when 
$\beta \agt 5$).  However, even at the lowest temperatures $T$ we always 
remain in the thermodynamic limit, i.e.~a large number of excited
states contribute to a possible symmetry breaking and  
the energy level spacing remains very small compared to $T$. 

\begin{figure}
\includegraphics[width=.47\textwidth]{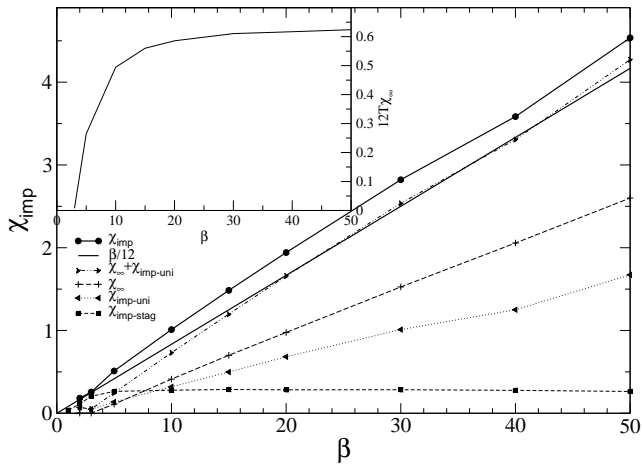}
\caption{The different contributions to the impurity susceptibility as a function of $\beta$. 
The error in the uniform impurity susceptibility is about $0.2$ because it is a difference of two large 
values. Inset: The ratio $12 T \chi_\infty$ as a function of $\beta$.}
\label{allimpsusc}
\end{figure}

The origin of a classical Curie impurity susceptibility from
a single vacancy in Eq.~(\ref{chiimp}) can intuitively
be understood by considering the classical limit of an Ising system with $N$ spins of size $S$ 
(here $S=\frac{1}{2}$).  Removing
one spin on the even sublattice at $(0,0)$ does not affect the long range order of such a 
system at all.  In that case
all the spins on the odd sublattice form one large effective spin of size $ NS/2$, 
while the spins on the even sublattice have total spin of $NS/2-S$.  The difference of these
two large spins will leave a net {\it classical} impurity spin of size $S$, which is free to rotate
toward an applied magnetic field.  
In the 2D Heisenberg antiferromagnet a similar mechanism can be invoked, 
except that the long range order 
amounts only to about 61\% of the classical limit\cite{Sandvik2,reger}. 
As can be seen in Fig.~(\ref{diagonal}) we can indeed find a field-induced longitudinal long range 
staggered order which approaches a limiting constant $\chi_\infty = \chi_{\rm stag}(\infty)$.
The value of $\chi_\infty$ approaches
a Curie behavior of about $0.6 \frac{S^2}{3k_BT}$ as can be seen in 
the inset of Fig.~(\ref{allimpsusc}), 
consistent with the fact that only about 60\% of the classically possible long range order 
couples to the field.  The corresponding contribution to the impurity susceptibility is given by 
\begin{equation}
\chi_{\rm imp-lr} = \sum_\mathbf{r\neq 0} (-1)^{x+y+1} \chi_\infty = \chi_\infty 
\sim 0.6 \frac{S^2}{3k_BT}.
\label{longrange}
\end{equation}
Since this contribution from long range order only amounts to 60\% of the 
leading impurity susceptibility in Eq.~(\ref{chiimp}),
the remaining 40\% and subleading contributions of $\chi_{\rm imp}$
must arise from a more local disturbance of Knight shifts in the vicinity around the vacancy.



The uniform part of the Knight shifts 
is found to drop off very fast with the distance from the impurity
to a value $\chi_{\rm pure}$ which corresponds to
the bulk susceptibility in a system without impurities $\chi_0/L^2$.
This is shown in Fig.~\ref{unipl} for various temperatures as a function of distance.
Closer inspection shows that the behavior on distance is
not completely isotropic, since the enhancement is stronger along the lattice axes
(at integer distances $|\mathbf{r}| = 1,2,3$).
The sum of this uniform part gives a net contribution to the impurity susceptibility
\begin{equation}
\chi_{\rm imp-uni} = \sum_{r\neq 0} \chi_{\rm uni}(\mathbf{r})-\chi_0,
\label{imp-uni}
\end{equation}
which turns out to contain
the remaining 40\% of the classical Curie impurity susceptibility of $1/12T$ in Eq.~(\ref{chiimp}).
This is shown in Fig.~\ref{allimpsusc} in comparison to other contributions to the impurity
susceptibility, where $\chi_{\rm imp-uni}$ is seen to be proportional to $\beta$ and the sum
$\chi_{\infty}+ \chi_{\rm imp-uni}$ follows the predicted form of the leading Curie contribution
$1/12T$ closely.

Finally, there is a staggered pattern of Knight shifts localized around the 
impurity $\chi_{\rm stag}(\mathbf{r}) -\chi_\infty$ as shown in the inset of Fig.~\ref{diagonal}.  
Interestingly,
the dependence on temperature is very weak both in shape and magnitude.
The net contribution to the impurity susceptibility 
\begin{equation}
\chi_{\rm imp-stag} = \sum_{r\neq 0} (-1)^{x+y+1} (\chi_{\rm stag}(\mathbf{r})-\chi_\infty) 
\label{imp-stag}
\end{equation}
is also small and contributes only to higher orders of $\chi_{\rm imp}$
as shown in 
Fig.~\ref{allimpsusc} compared to the other contributions.  
Obviously $\chi_{\rm imp} = \chi_{\rm imp-stag} + \chi_{\rm imp-lr} + \chi_{\rm imp-uni}$.


We now consider the case of two impurities that are separated sufficiently 
far away from 
each other, so that any quantum interference 
can be neglected.  As long as the correlation length $\xi$ is smaller than the inter-impurity distance 
no interference can be found and the impurity contributions to the Knight shifts and the 
susceptibility simply add.  
In the low temperature limit the correlation length grows beyond the
system size and the staggered order extends through the entire system.  In this case
it makes a significant difference if the two vacancies are on the same or on different sublattices.
\begin{figure}
\includegraphics[width=.47\textwidth]{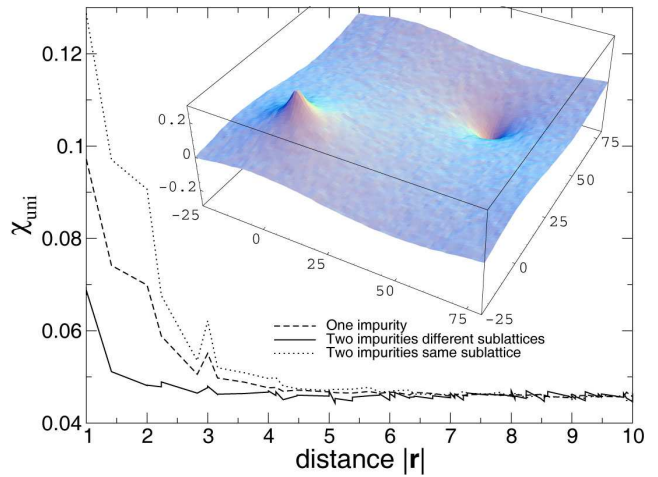}
\caption{ (Color online) The uniform part of the Knight shifts around an impurity at $(0,0)$ 
as a function of distance at $\beta=20$.  For a second impurity on the same sublattice $(48,48)$ (dotted)
we find twice the uniform enhancement compared to the single impurity case (dashed).  For a 
second impurity on the other sublattice $(48,49)$ (solid) there is a strong reduction. Inset: Staggered
part for two impurities on different sublattices.}
\label{twoimpuni}
\end{figure}


For two vacancies on {\it different} sublattices 
the field-induced long range order cancels exactly, giving no contribution to the 
impurity susceptibility.  
Remarkably, the uniform contribution is also strongly affected 
as shown in Fig.~\ref{twoimpuni} for impurities
located at $(0,0)$ and $(48,49)$ in comparison to the uniform Knight shift data of just one
impurity. Even though the impurities are located 
a distance of $|\mathbf{r}| \sim 69$ sites apart, we find a strong reduction of the uniform Knight shifts 
within just a few lattice sites around each impurity in Fig.~\ref{twoimpuni}.  
This reduction goes far beyond simple 
interference and must be taken as a collective effect of the entire system. 
The resulting impurity contribution $\chi_{\rm imp-uni}$ from 
the uniform Knight shifts in Eq.~(\ref{imp-uni}) is small and has no Curie-like temperature behavior, 
i.e.~only the subleading corrections remain.
On the other hand, the {\it local staggered} enhancement 
of the Knight shifts is remarkably unaffected by interference effects, but has the 
opposite phase around each impurity as shown in the inset
of Fig.~\ref{twoimpuni}.  Therefore, the sum in Eq.~(\ref{imp-stag}) gives exactly twice the magnitude of 
$\chi_{\rm imp-stag}$ compared 
to one single impurity (the opposite phase
cancels with the minus sign from the different sublattice in Eq.~(\ref{imp-stag})).  
This part $\chi_{\rm imp-stag}$ now becomes the main
contribution to the impurity susceptibility, but has no divergent Curie-like behavior as
discussed above.  
This is in agreement with findings in Ref.~[\onlinecite{Sandvik1}] that the 
impurity susceptibility of two impurities on different sublattices corresponds to 
subleading terms only.

For two vacancies on the {\it same} sublattice 
we find a doubling of both the induced longitudinal long range order
$2 \chi_{\infty}$ and the uniform enhancement as shown in Fig.~\ref{twoimpuni}.   
Since two vacancy sites contribute in Eqs.~(\ref{longrange}) and (\ref{imp-uni})
this implies that the Curie-like impurity susceptibilities $\chi_{\rm imp-lr}$ and $\chi_{\rm imp-uni}$
are {\it quadrupled} as if two virtual
spins form a triplet.  On the other hand, the local staggered enhancement of 
Knight shifts close to each impurity again remains almost completely unaffected, compared to the single 
impurity case in the inset of Fig.~\ref{diagonal}.
The Curie-like contributions $\chi_{\rm imp-lr}$ and $\chi_{\rm imp-uni}$  therefore increase 
{\it quadratically} with the number of same-sublattice impurities $N$, while
the subleading contribution $\chi_{\rm imp-stag}$
increases linearly with $N$ {independent} of sublattice 
$\chi_{\rm imp} \sim N^2/12T + N {\cal O}\left(\ln(C/T)\right)$.

In the case of two impurities that are very close to each other, 
quantum effects are bound to also play a role and 
a comparison with the interference of two 
single impurities is not useful. For two neighboring impurities we find that also subleading 
contributions to the impurity susceptibility are smaller\cite{Sandvik1}.

In an infinite system with 
a small random impurity concentration $\rho$ the impurities act independently
at high and intermediate temperatures (assuming an average distance of at least several lattice sites).
As the temperature is lowered, however, vacancies within the distance of the exponentially growing
correlation length $\xi$ start to feel each other,  i.e.~the effective number of correlated impurities
within an antiferromagnetic domain is $N\propto \xi^2 \rho $.
The subleading contribution from the local staggered part $\chi_{\rm imp-stag}$ always 
adds independent of sublattice and will therefore be 
of order $N$, but is not divergent with temperature.
For the Curie-like contributions $\chi_{\rm imp-lr}$ and $\chi_{\rm imp-uni}$
the difference of impurities on the two sublattices 
plays an important role, which is of order $|N_A-N_B| \sim \sqrt{N}$\cite{Sandvik2}.  
Our simulations show that the induced long-range order is proportional to the excess of impurities
on one sublattice $|N_A-N_B| \chi_\infty \propto \xi \sqrt{\rho} \chi_\infty$ 
while the uniform enhancement is proportional to 
$\pm (N_A-N_B)$ on the sublattice A/B up to subleading terms, i.e.{\it negative} on the minority sublattice.
The contributions $\chi_{\rm imp-lr}$ and $\chi_{\rm imp-uni}$
in the sums (\ref{longrange}) and (\ref{imp-uni}) 
therefore both increase quadratically with the difference $(N_A-N_B)^2 \sim N$.
Hence we predict that a Curie susceptibility of $\rho S^2/3 k_B T$ survives 
even in the thermodynamic limit,
contrary to conjectures made in Ref.~[\onlinecite{Sandvik2}].  In NMR experiments induced staggered
Knight shifts of order $\xi \sqrt{\rho} \chi_{\infty} B$ throughout the lattice should be 
directly observable.   

In conclusion we have analyzed the local response around vacancies 
in the 2D Heisenberg antiferromagnet in the low temperature limit.
These Knight shifts can be analyzed in terms of staggered and uniform 
contributions, which are mostly isotropic on the 2D lattice.  
In the presence of a small magnetic field a single vacancy induces a staggered 
pattern of Knight shifts over the entire lattice in the low temperature limit, giving a
longitudinal N\'eel order $\chi_\infty \sim 0.6 /12 T$. 
This field-induced order corresponds to a free classical moment
accounting for 60\% of the Curie impurity susceptibility.  The remaining 40\% of the 
magnetic moment is found to be located in
a uniform enhancement of Knight shifts in the immediate neighborhood of the vacancy.  
Additionally a staggered pattern in a finite range around the vacancy is also found,
which turns out to have a largely temperature independent, universal shape and gives rise to subleading 
corrections of $\chi_{\rm imp}$.

When more than one vacancy is introduced in the low temperature limit, 
the Curie terms in the impurity susceptibility cancel if
the vacancies are on different sublattices with a remarkable long-distance reduction of the
uniform Knight shifts in the close vicinity of each vacancy.  
For vacancies on the same sublattice both the 
induced long range order and the uniform part increase, leading to a Curie-like impurity 
susceptibility that increases {\it quadratically} with the number of impurities.  Both uniform and 
long range staggered parts appear to be contributing to virtual classical impurity spins that are
ferromagnetically/antiferromagnetically linked for impurities on the same/opposite sublattice.
The local staggered enhancement around each impurity remains unaffected by interference and temperature
in any case, contributing to subleading corrections of $\chi_{\rm imp}$. 
In the thermodynamic limit cancellation and non-linear enhancement effects even out 
from statistical arguments, leaving a Curie susceptibility  of $\rho /12 T$. 

Some of the results can also be generalized for other quantum 
antiferromagnets, like Heisenberg models 
in higher dimensions and/or with higher spin.  In particular, a vacancy 
will always induce a longitudinal antiferromagnetic response $\chi_\infty$
over the entire lattice, the magnitude of which
is linked to the 
order parameter as $T\to 0$ (80\%  in Eq.~(\ref{longrange}) for a 3D spin-1/2 model). 
The remaining impurity susceptibility in Eq.~(\ref{chiimp}) must be found in a uniform enhancement around
the vacancies. The virtual classical impurity spins are antiferromagnetically/ferromagnetically linked 
depending on the sublattice as seen above.  
In the spin-1/2 chain no long-range order and no Curie impurity susceptibility is 
found.  However, the induced Knight shifts increase with a distance from 
the vacancy\cite{eggert} and 
boundary terms give a Curie contribution that is logarithmically reduced for low 
temperatures from non-universal boundary effects\cite{satoshi}.
For vacancies in quantum antiferromagnets with a gap, 
a more local antiferromagnetic pattern is induced\cite{Sandvik2,Martins}, 
leading to a Curie susceptibility corresponding 
to  virtual quantum spins, which are not linked to each other.  The effect of substitutional impurities
with higher spin is much less clear, although a coupling to the antiferromagnetic order 
can again be expected.

\acknowledgments
We are very thankful for helpful discussions with Mats Granath, Holger Frahm, Henrik Johannesson,
and Matthias Vojta.  This project was in part supported by the Swedish Research Council.
Computer time was allocated through Swegrid grant SNIC 011/04-15.

\end{document}